\documentstyle{article}

\newcommand{\mathbf}{\bf}

\begin{document}

\begin{center}
{\huge\bf  The Edge Currents and Edge Potentials in IQHE }
\end{center}

\vspace{1cm}
\begin{center}
{\large\bf
F.GHABOUSSI}\\
\end{center}

\begin{center}
\begin{minipage}{8cm}
Department of Physics, University of Konstanz\\
P.O. Box 5560, D 78434 Konstanz, Germany\\
E-mail: ghabousi@kaluza.physik.uni-konstanz.de
\end{minipage}
\end{center}

\vspace{1cm}

\begin{center}
{\large{\bf Abstract}}
\end{center}

\begin{center}
\begin{minipage}{12cm}
It is shown that an observed length in the potential drops across  
IQHE samples  is a universal length for a given value of magnetic  
field which results from the quantum mechanical uncertainty  
relation.

\end{minipage}
\end{center}

\newpage

We showed recently that the microscopic theory of IQHE \cite{all}  
can be given by the canonical quantization of a semi-classical  
theory of the "classical" Hall-effect CHE \cite{mein}.

The action functional for this is the semi-classical  
Schroedinger-Chern-Simons action for a 2-D non-interacting carrier  
system with the usual minimal electromagnetic coupling on a   
2+1-dimensional manifold $M = \Sigma\times\mathbf R$ with spatial  
boundary. We showed also that the constraints of the theory forces  
the coupled electromagnetic potential to be an almost pure gauge  
potential, i. e. with an almost vanishing field strength and they  
forces also the potential to exist only on the edges \cite{mein}.  
Thus, according to our model we have to do in IQHE case with an  
almost pure "edge" gauge potential \cite {almost}. Accordingly, in  
view of the Ohm's equations the edge currents are the prefered  
currents under these constraints.

Here we show that the recent results on the potential drops across  
IQHE samples near the edges \cite{DK} follow the universal  
uncertainty relation of quantum mechanics, in view of the  
universality of the QHE.

To begin, recall that there are two fundamental aspects of  
potential which has to be considered:

1) that the potential itself is non-observable but some functions  
of it becomes observable.

2) that a pure gauge potential is according to the quantum  
mechanics a non-vanishing quantity (see below).

Let us first explain the situation from the more fundamental point  
of view of quantum mechanics.

\medskip
For a charged system, e. g. electrons in magnetic fields, the  
energy uncertainty is given by the minimum amount of energy, i. e.  
the ground state energy. This amount of energy is proportional to  
the applied magnetic field strength. On the other hand, an energy  
uncertainty is correlated with a position uncertainty for electrons.  
Thus, quantum mechanically there is always an uncertainty of  
position of the electronic currents on the surface which is related  
with the width of the electron orbit.
Therefore, if we consider the uncertainty of momentum equal to  
$(2m_e\; \Delta E)^{\frac{1}{2}}$ with $\Delta E = E_{n+1} - E_{n} =  
{\displaystyle{\frac{\hbar \omega_c}{2}}}$ and $\omega_c =  
{\displaystyle{\frac{eB}{m_e}}}$, then the mentioned position  
uncertainty in $2-D$ is given by $\Delta Y = \Delta X =  
({\displaystyle{\frac{\hbar}{eB}}})^{\frac{1}{2}}$ which is the  
magnetic length $l_B$. Since, the edge current is defined as the  
current which flows, in the ideal case, close to the edge within the  
length scale of the magnetic length \cite{kk}. This means that one  
should expect that according to  Ohm's equations for QHE, in the  
ideal case, also the potential distribution on the sample should be  
close to the boundary of sample within a distance which is  
proportional to the magnetic length.

Furthermore, one must take into account that despite of classical  
physics in quantum physics there are relevant quantities which are  
prevented to become zero in view of the uncertainty relations. To  
these relevant quantities in the QHE case it belongs the  
electromagnetic potential and its field strength. Equivalently, in  
quantum mechanics only global quantities $\oint A_m dx^m = \int\int  
B ds$ are relevant but not their local components $A_m$ or $B$.

Thus, if we consider for example $\Delta P_y = \Delta A_y = e A_y$,  
then there is an uncertainty relation $e A_y \cdot \Delta Y = e A_y  
\cdot l_B = \hbar$.

Accordingly, a pure gauge potential which should be zero  
classically for example within an IQHE sample, is however quantum  
mechanically non-zero and has a value of $\Delta A_y =  
\displaystyle{\frac{\hbar}{e l_B}}$.

\medskip
In view of the relations between the magnetic field strength $B$,  
magnetic length and the global density of electrons $n$ with the  
filling factor $\nu$, i. e. $l^2_B =  
{\displaystyle{\frac{\hbar}{eB}}} = {\displaystyle{\frac{\nu}{2\pi  
n}}}$, it is obvious that a variation of only one of these factors  
changes the magnetic length and so it changes also the current  
position and the potential distribution on the sample. However, if  
$B$ or ${\displaystyle{\frac{\nu}{n}}}$ remain the same for various  
IQHE samples, then the magnetic length should be invariant for all  
these samples under the IQHE conditions independent of their  
geometries and other factors.

These are the quantum theoretical basics of what is observed in the  
mentioned experiments for the potential drops \cite{DK}, where the  
authors report that they observed potential drops across the  
IQHE-samples over a length of $100 \mu m$ from the edge of samples.  
We show that this length which has the {\it magnitude} of  
$|l_B^{-1}|$ for the given data in Ref. \cite{DK} is indeed a  
universal quantity for a given $B$ or for a given  
${\displaystyle{\frac{\nu}{n}}}$ \cite {nn}.

\medskip
Furthermore, as we mentioned above the electromagnetic potential is  
in view of its gauge dependence non-observable. The observables  
related with the potential or those related with its field strength  
are phase angles given by the closed path integral of potential or  
the surface integral of field strength, which are observable by the  
quantum mechanical interfrence patterns. Equivalently, a constant  
potential multiplied by a proper length, e. g. by the circumference  
of mentioned closed path is also observable. For example according  
to the definition of magnetic length $l^{2}_B =  
{\displaystyle{\frac{\hbar}{e B}}}$ we have \cite{gauge} (see also  
below):

\begin{equation}
l^{2}_B B = l_B A = \frac{\hbar}{e}\;\;\;,
\end{equation}
\label{m lang}

which is equivalent to the definition of magnetic flux quantum  
through $\int\int B ds = \oint A_m dx^m =  
{\displaystyle{\frac{h}{e}}}$, where the potential component $A$ in  
(1) is the relevant component of electromagnetic gauge potential  
according to the $A_m = B.x_n \epsilon_{mn}$ gauge \cite{gauge}.

\medskip
Recall also that, the potential and magnetic field have always  
dimension $L^{-1}$ and $L^{-2}$ respectively. Thus, it is natural  
that under the IQHE conditions where $\sigma_H =  
\displaystyle{\frac{ne}{B}}$ is quantized according to $\sigma_H =  
\nu \displaystyle{\frac{e^2}{h}}$, one obtains a purely geometrical  
relation between potential and magnetic length.

\medskip
Moreover, as a general result let us mention that, if one considers  
the relation (1) in form $2 \pi l_B A = 2 \pi l^{2}_B B =  
{\displaystyle{\frac{h}{e}}}$ as given according to the flux  
quantization for electrons flow in the IQHE edge current on a ring  
with radius and width both equal to $l_B$. Then, one obtains with  
the given $l_B$ according to the data in Ref. \cite{DK} for $ A =  
{\displaystyle{\frac{\hbar}{e}}}l^{-1}_B$ a value about $100 \mu m$  
for $A$, which is the mentioned observed legth for potential drops  
\cite{DK} \cite{dim}.

This result show that in view of the definition of magnetic length  
the measured value of $100 \mu m$ is a fundamental value for IQHE  
experiments on those samples independent of other sample parameters.  
It shows also that for a pure "edge potential" $A$ which should  
exists classically exactly on the edges of sample and it should be  
zero in the rest of sample \cite{mein}, we have however $\Delta A  
\neq 0$ in view of $\Delta A \Delta X =  
\displaystyle{\frac{\hbar}{e}}$. It is also in view of the $L^{-1}$  
dimension of $A$ that one obtains $\Delta A \neq 0$ within a width  
of $l_B ^{-1}$.

\medskip
Furthermore, the relation between edge current and the above  
discussed "edge potential" or the edge potential drops should be  
understood in the following way with respect to the above  
considerations:

Obviously, $B \cdot x^m = \epsilon^{mn} A_n$ is a solution of the  
Ohm's equations $j_m = \epsilon^{mn} \sigma_H E_n$ with $j^m = ne  
\displaystyle{\frac{dx^m}{dt}}$ and $E_n =  
\displaystyle{\frac{dA_n}{dt}}$, if we use as usual $\sigma_H =  
\displaystyle{\frac{ne}{B}}$ in the quantum Hall limit: $\omega_c  
\tau \gg 1$. From this solution it results that the edge current  
$j_m$ should flow within a width of $\Delta X = l_B$ in view of the  
already mentioned relation $B \cdot \Delta X = \Delta A$, in  
agreement with its definition. Moreover, one can prove directly the  
measured value of potential drops in Ref. \cite{DK} from the  
relation $B \cdot \Delta X = \Delta A$.
Thus, for an applied magnetic field $B$ about $1 \; Tesla$ and for  
the $l_B$ value which is known to be $10^{-2} \mu m$ from the  
$\displaystyle{\frac{\nu}{2\pi n}}$ value of the given sample, one  
obtains according to $B \cdot \Delta X = \Delta A$ for $\Delta A$ a  
vlue of  $100 \mu m$.

In this way the edge current of charged carriers which flows within  
a width of $l_B$ causes a potential drop of the measured width. The  
same calculation should be done for the experiments with filling  
factor $\nu = 4$ about which it is reported in Ref. \cite {font}.  
The theoretical result agrees also in this case with the measured  
result. Moreover, the same results can be obtained according to the  
relation $\displaystyle{\frac{\hbar}{e}} B = A ^2$
in view of the definition of $l_B$. For a given value of $B$ which  
corresponds with $\nu$ according to the relation $\nu =  
\displaystyle{\frac{n h}{eB}}$ this is an invariant relation for the  
potential $A$ which shows the general validity of the experimental  
results and also of our theoretical result.

\medskip
To be precize, let us mention that in other experiments  
\cite{font}, where the electronic concentration is almost the same  
as in Ref. \cite{DK} but the filling factor is $\nu =4$, one  
observed potential drops of $\approx 70 \mu m$. This is in good  
agreement with our theoretical result, since for $\nu =4$ filling  
factor one obtains according to the data of Ref. \cite{font} a  
magnetic length $l^{\prime}_B \approx 1.4 l_B \approx 1.4 \cdot  
10^{-2}$ $\mu m$, where $l_B \approx 10^{-2}$ $\mu m$ is the  
magnetic length of samples in Ref. \cite{DK}.
Thus, the theoretical value of $ A =  
{\displaystyle{\frac{\hbar}{e}}}(l^{\prime}_B)^{-1}$ becomes  
$\approx 70 \mu m$ which is indeed the measured value according to  
Ref. \cite{font} (see also \cite{dim}).

\medskip
This circumstance explains why one observes potential drops within  
such a distances from the edges of the IQHE samples \cite{DK}  
\cite{font}.

\medskip
Therefore, one should claim that the measured penetration length of  
electromagnetic potential on IQHE samples should depend, according  
to the theoretical value of $ A =  
{\displaystyle{\frac{\hbar}{e}}}(l_B)^{-1}$, only on the related  
value of $l^{-1}_B$ \cite{dim}.

\medskip
>From theoretical point of view the origin of these empirical  
results should lie, as it is mentioned already, in the quantum  
mechanical uncertainty- principle, where a charged particle in  
presence of magnetic fields acquires a position uncertainty $\Delta  
Y = \Delta X = l_B$.
Thus, considering $\Delta P = \Delta A = e A$, we are given under  
quantum mechanical conditions of QHE, the uncertainty relation $  
A\cdot l_B = \displaystyle{\frac{\hbar}{e}}$. Here  
${\displaystyle{\frac{\hbar}{e}}}$ plays the same role in the  
quantum electrodynamical uncertainty as that played by $\hbar$ in  
the quantum mechanical uncertainty.

Therefore, in view of the fact that the value of  
${\displaystyle{\frac{\hbar}{e}}}$ is a  fixed quantity, the value  
of potential (drop) under IQHE conditions is always given by $A =  
{\displaystyle{\frac{\hbar}{e l_B}}}$, as it is confirmed by results  
in Ref. \cite {DK} and \cite{font}, no matter what other relevant  
quantities are.

Thus, in any IQHE sample one should measure for the potential drops  
on the edges the {\it related} value of $A =  
{\displaystyle{\frac{\hbar}{e l_B}}}$ according to the value of  
$l_B$ from the experimental data of sample.

\medskip
In view of the fact that this is a result from the uncertainty  
principle and as such it is an invariant result, it depends only on  
the basics of "magnetic" quantization, i. e. on the uncertainty  
principle in quantum electrodynamics.

\medskip
Furthermore, it is expected that the observed length of the  
potential drop should be related with parameters of samples. This is  
indeed true, if one recalls that the concentration of charge  
carriers is indeed the main parameter of the sample and also the  
magnetic length depends on it.

\medskip
In conclusion let us mention that such a penetration length is also  
comparable with London's penetration length in superconductivity  
\cite{sup}.

\bigskip
Footnotes and references


\begin{thebibliography}{100}

\bibitem{all}
For a general review on QHE and its experimental setting see:

[1a] R.E. Prange and S.M. Girvin, ed., The quantum Hall effect,  
Graduate Texts in Contemporary Physics (Springer, New York, 1987);

[1b] A.H. Macdonald, ed., Quantum Hall effect: A Perspective,  
Perspectives in Condensed Matter Physics (Kluver Academic  
Publishers, 1989)

[1c] G. Morandi, The role of Topology in Classical and Quantum  
Physics, Lecture Notes in Physics m7 (Springer, New York 1992)

[1d] M. Janssen, et al, ed., J. Hajdu, Introduction to the Theory  
of the Integer Quantum Hall effect (VCH-verlag, Weinheim, New York,  
1994)

[1e] V. J. Emery (editor), Correlated Electron Systems, (World  
Scientific, Singapore 1993)

[1f] J. Froehlich, T. Kerler, Nuc. Phys. B354 (1991) 369-417.

[1g] A. Shapere and F. Wilczek (editors): Geometric Phases in  
Physics ( World Scientific, Singapore, 1989)


\bibitem{mein}
F. Ghaboussi, "On the Integer Quantum Hall Effect", KN-UNI-preprint-95-1;
"A Model of the Integer Quantum Hall Effect", KN-UNI-preprint-95-2,  
submitted for publication; See also "On the Hall-Effect and its  
Quantization", KN-UNI-preprint-95-3, submitted for publication.


\bibitem{almost}
A pure gauge potential in QHE case is a potential in a multiply  
connected region, thus it can not be gauged away everywhere in the  
region. Its field strength is in some parts of the region zero and  
in some other parts not. Quantum mechanically one can observe the  
influence of such pure gauge potential also in those regions where  
the field strength is zero (see Bohm-Aharonov effect).

\bibitem{kk}
K. von Klitzing, Physica B 204 (1995) 111-116;
R. Knott, W. Dietsche, K. von Klitzing, K. Eberl and K. Ploog,  
Semicond. Sci. Technol. 10 (1995) 117-126

\bibitem{DK}
W. Dietsche, K. v. Klitzing and K. Ploog, Potential Drops Across  
Quantum Hall Effect Samples- In the Bulk or Near the Edges? MPI fuer  
Festkoerperforschumg Stuttgart-preprint 1995;

\bibitem{nn}
According to the data about the IQHE samples in Ref. \cite{DK} the  
global concentration is

$n = 3.7 \cdot 10^{11} cm^{-2}$ and $\nu = 2$. Thus, one obtains  
$l_B \approx 10^{-2} \mu m$.

The measured pentration length is given to be about $100 \mu$ m  
which is almost exactly $|l^{-1}_B| \mu m$.

\bibitem{gauge}
In view of the gauge invariance of electrodynamics one is free to  
choose relevant gauges to retain the true degrees of freedom of the  
electromagnetic field. We use here for the homogeneous magnetic  
field $B$ the so called Landau gauge $ A = A_y = B \cdot x$ for the  
only relevant component of potentential.

Recall further that the uncertainty relation discussed above, i. e.  
$e \Delta A_y \Delta Y = \hbar$ together with $e \Delta A_y = B  
\cdot \Delta X$ result in the flux quantization relation $B \cdot  
\Delta X \Delta Y = \displaystyle{\frac{\hbar}{e}}$

\bibitem{dim}
Recall that the measured width of the potential drops should be  
considered theoretically according to the dimensinal structure where  
$\hbar$ contains $L^2$ dimensions according to its definition.


\bibitem{font}
P. F. Fontein, et al., Phys. Rev. B., 43, 12090 (1991). The given  
data in this report which are relevant for our calculation are $n =  
5.0 \cdot 10^{15}\, m^{-2}$ and $\nu = 4$.

\bibitem{sup}
It is well known that superconducting effects can be considered as  
to be related with the QHE: see Ref. [1f]; R. B. Laughlin: in Ref.  
[1g]; and A. Karlhede, et al.: in Ref. [1e]; See further for  
empirical confirmations: D. Jerome, in J. G. Bednorz, K. A. Mueller  
(Eds), Superconductivity,

( Spriger-Verlag, Berlin 1990).
\end{thebibliography}
\end{document}